\documentclass[12pt]{article}

\usepackage{scicite}
\usepackage{graphicx}
\usepackage{times}

\usepackage{hyperref}

\newenvironment{sciabstract}{%
\begin{quote} \bf}
{\end{quote}}

\title{Two-dimensional few-atom noble gas clusters in a graphene sandwich}

\author
{Manuel Längle$^{1,2\ast}$, Kenichiro Mizohata$^{3}$, Clemens Mangler$^{1}$,\\
Alberto Trentino$^{1,2}$, Kimmo Mustonen$^{1}$, E. Harriet {\AA}hlgren$^{1}$,\\
Jani Kotakoski$^{1\dagger}$\\
\footnotesize{$^{1}$University of Vienna, Faculty of Physics, Boltzmanngasse 5, 1090 Vienna, Austria}\\
\footnotesize{$^{2}$University of Vienna, Vienna Doctoral School in Physics, Boltzmanngasse 5, 1090 Vienna, Austria}\\
\footnotesize{$^{3}$University of Helsinki, Department of Physics, P.O. Box 43, FI-00014 Finland}\\
\footnotesize{$^\ast$E-mail: manuel.laengle@univie.ac.at}
\footnotesize{$^\dagger$E-mail: jani.kotakoski@univie.ac.at}
}

\date{}

\begin{document}


\baselineskip24pt


\maketitle


\begin{sciabstract}
    Van der Waals atomic solids of noble gases on metals at cryogenic temperatures were the first experimental examples of two-dimensional systems.
    Recently such structures have also been created on surfaces under encapsulation by graphene, allowing studies at elevated temperatures through scanning tunneling microscopy.
    However, for this technique, the encapsulation layer often obscures the actual arrangement of the noble gas atoms.
    Here, we create Kr and Xe clusters in between two suspended graphene layers, and uncover their atomic structure through direct imaging  with transmission electron microscopy.
    We show that small crystals ($N<9$) arrange as expected based on the simple non-directional van der Waals interaction.
    Crystals larger than this show some deviations for the outermost atoms, possibly enabled by deformations in the encapsulating graphene lattice.
    We further discuss the dynamics of the clusters within the graphene sandwich, and show that while all Xe clusters with up to $N \sim 100$ remain solid, Kr clusters with already $N \sim 16$ turn occasionally fluid under our experimental conditions with an estimated pressure of ca. 0.3~GPa.
    This study opens a way for the so-far unexplored frontier of encapsulated two-dimensional van der Waals solids with exciting possibilities for fundamental condensed matter physics research and possible applications in quantum information technology.
\end{sciabstract}

Two-dimensional (2D) van der Waals atomic crystals are one of the simplest conceivable solids, described to a large extent with the simple pairwise interaction approximated by the Lennard-Jones potential.
Such structures formed by noble gas atoms have been experimentally studied since the 1960's at cryogenic temperatures~\cite{jortner_localized_1965}.
Since the van der Waals interaction between noble gas atoms is isotropic, and only depends on the inter-atomic distance, the 2D crystals assume a close-packed hexagonal structure. This was confirmed through ultralow-energy electron diffraction experiments already in 1975~\cite{cohen_xe_1976}.
First real space images of 2D noble gas crystals were recorded via scanning tunneling microscopy in 1998 on a graphite surface~\cite{bauerle_studies_1998}. 

Studies on van der Waals structures were for a long time limited to cryogenic temperatures due to the weakness of the van der Waals interaction.
Only after the discovery of the impermeability of graphene to the smallest gases in 2008~\cite{bunch_impermeable_2008}, it became possible to create such structures at elevated temperatures by trapping noble gases within the van der Waals gap between graphene and the supporting surface.
Such Xe clusters have already been imaged using scanning probe microscopy~\cite{valerius_annealing_2017, herbig_graphene_2016}.
However, the graphene encapsulation complicates imaging the atomic structure of the crystallites which has even lead to misidentification of such structures in the past~\cite{herbig_interfacial_2014,herbig_comment_2015}.
It has been also shown that Kr can be implanted under the first graphite layer, essentially into the interface between graphene and bulk graphite~\cite{yoo_growth_2018}.
Implantation deeper into graphite should also be possible, but the rigidity of the thicker graphite layers on both sides of the implanted species (as compared to a single layer) might make it impossible to form solid structures without considerable damage to the surrounding lattice.

Here, we create 2D few-atom noble gas clusters by implanting Kr and Xe between suspended graphene sheets using ultralow-energy ion irradiation, and image them at room temperature using atomic-resolution scanning transmission electron microscopy (STEM), complemented by electron energy loss spectroscopy for confirming the chemical identity of the atoms. 
Our samples consist of both mechanically exfoliated bi- and few-layer graphene as well as commercial graphene grown via chemical vapor deposition (CVD).
For the CVD-grown samples, two graphene monolayers were stacked on top of each other to create an artificial double layer with a random misorientation between the graphene crystals.
This allows us to compare samples where the van der Waals gap has never been exposed to air or contamination to samples where the two layers are joined under ambient conditions, and different moiré periodicities.
In all cases, similar noble gas clusters were found with no discernible difference.
Samples were irradiated with singly charged ultralow-energy Kr or Xe ions to introduce the noble gases.
For Kr-irradiated samples, successful implantation between two graphene layers was achieved at 60~eV, and for Xe-irradiated samples at 55~eV and 65~eV, close to the simulated lower bound for Xe implantation of 70~eV~\cite{shiryaev_behavior_2021}.
One additional sample was implanted using 30-eV Xe ions at the University of Vienna using a plasma source within the vacuum system~\cite{mangler_materials_2022} connected to the Nion UltraSTEM 100 microscope used here for imaging.
These energies are close to the 25~eV that has been shown to allow implanting He, Ne and Ar into the van der Waals gaps of graphene-covered metal surfaces~\cite{villarreal_breakdown_2021}. 

After the implantation, the samples were imaged via STEM at 60~kV using either the high or the medium angle annular dark field (HAADF and MAADF) detectors.
Before being inserted into the vacuum system connected to the microscope, the samples were baked in vacuum at ca. 150$^\circ$C for at least 8~h.
Some samples were additionally cleaned before imaging with a laser at the microscope column~\cite{tripathi_cleaning_2017}.
Larger scale images of implanted samples with many clearly visible clusters are shown in Fig.~S1 and Fig.~S2 along with the recorded spectroscopic signatures confirming the elemental identification as Kr and Xe, respectively.

\begin{figure}[h!]	
	\centering
	\includegraphics[width=1.0\textwidth]{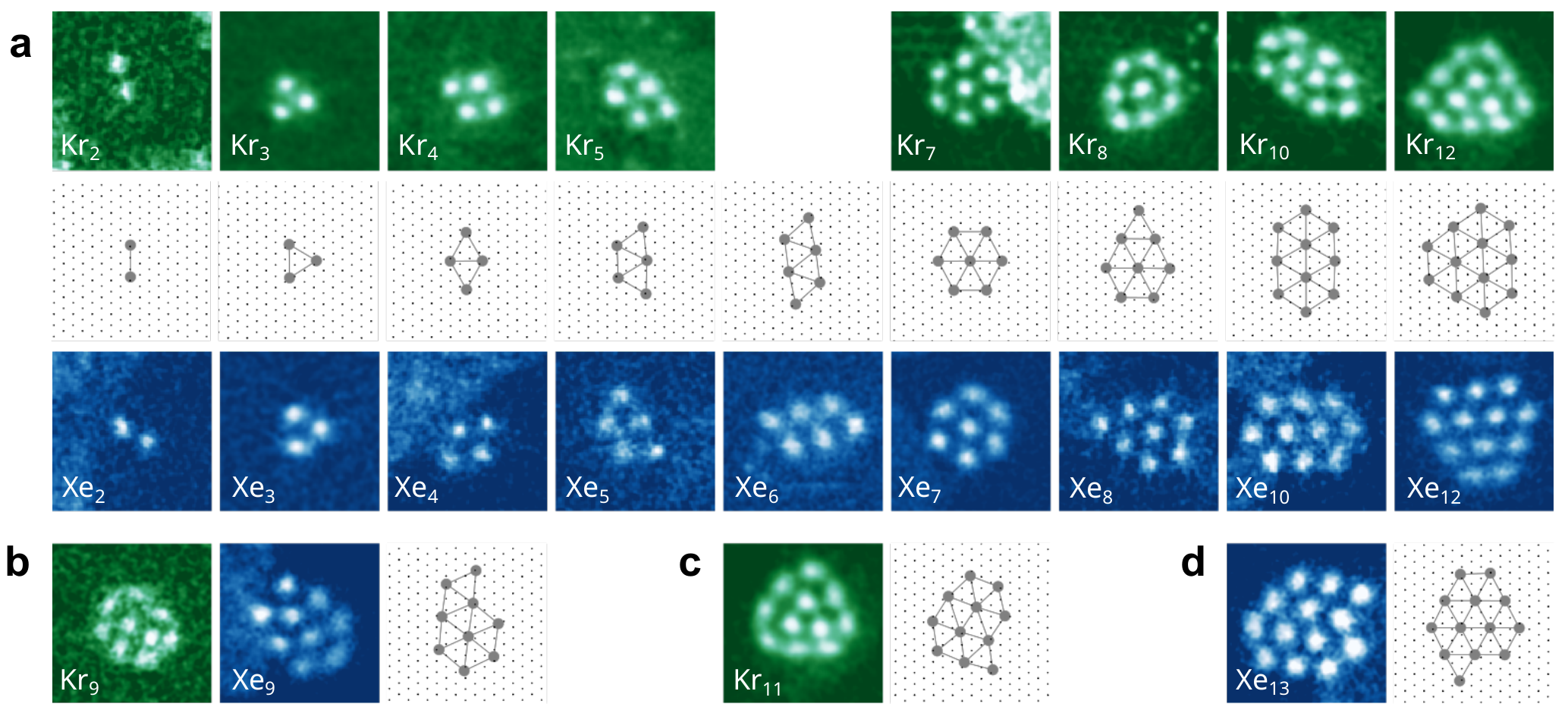}
	\caption{
        {\bf Structure of small clusters.}
        (a) Filtered STEM-ADF images (see Methods) of Kr (top row) and Xe (bottom row) clusters with $N \in [2, 12]$, including all cases where the experimentally observed structure corresponds to that with the lowest energy according to simulations. 
        The corresponding simulated structures after energy optimization are shown in the middle row for Xe (Kr has the same structure, see Fig.~S7 for all simulated structures).
        Carbon atoms are shown with small gray dots, and Xe atoms with large dots connected with lines.
        (b) Example STEM-ADF images of nine-atom Kr and Xe clusters along with the simulated lowest-energy configuration for Kr.
        (c) Example STEM-ADF image of a eleven-atom Kr cluster along with the simulated lowest-energy configuration for Kr.
        (d) Example STEM-ADF image of a thirteen-atom Xe cluster and the corresponding lowest-energy structure for Xe.
        All images have a size of $2\times 2$~nm$^2$.
    }
	\label{fig:crystals}
\end{figure}

Due to their significantly higher atomic number as compared to the carbon atoms of graphene, both Kr and Xe are easily visible in clean areas of the sample in the $Z$-contrast~\cite{krivanek_atom-by-atom_2010} STEM images.
Clusters with $N_\mathrm{Kr} \in [2, 12]$ and $N_\mathrm{Xe} \in [2, 13]$ atoms are shown in Fig.~\ref{fig:crystals}a-d along with the simulated energy-optimized structures.
All clusters appear flat, and assume a hexagonal close-packed structure.
The same cluster shapes are found both for Xe as well as Kr, and in samples using both exfoliated graphene and graphene grown via CVD.
Therefore, there is no plausible way that the clusters could contain any other element that remains invisible in our microscopy images.

The hexagonal close-packed structure minimizes the number of lower-coordinated atoms while at the same time maximizing the number of nearest neighbors within the cluster (see also Fig.~S3 in the supplementary material), demonstrating that the interaction between the atoms within each cluster is strong enough to dominate its shape under these conditions.
Up to $N<9$, all clusters in the experiment are found also in the optimal structure as predicted by the simulations, although in some cases also other structures with same $N$ are observed (see Fig.~S4).
Most notably, the Kr$_9$ cluster also appears in a shape where the atoms are arranged along two lines.
In this unexpected Kr$_9$ structure the number of van der Waals bonds with approximately the nearest neighbor distance ($d_{nn}$) is only 14 (in one case even 13), whereas in the ideal structure it is 17.
In addition to this specific case of Kr$_9$, the atoms in the outer rim of clusters with $N\geq 9$ appear to have less defined positions (also seen in Fig.~\ref{fig:crystals}b \& c).
These deviations from the ideal structure may arise from factors other than the interaction between the noble gas atoms (for example from deformations in the encapsulating graphene structure, as discussed below).
We also point out that one Kr$_{10}$ configuration, as well as Kr$_{11,12}$ all show the same cluster with occasional detachment and attachment of one or two atoms (the image series is shown in Fig.~S5).
In many cases, we also observe shape changes both in the simulations (Fig.~S6), as well as in the experimental images (Fig.~S4).
Although we did not find Kr$_6$, Xe$_{11}$ or Kr$_{13}$ within our images of small crystals, there is no reason to assume that they would be less stable than the clusters shown in Fig.~\ref{fig:crystals}.

\begin{figure}[h!]	
	\centering
	\includegraphics[width=1\textwidth]{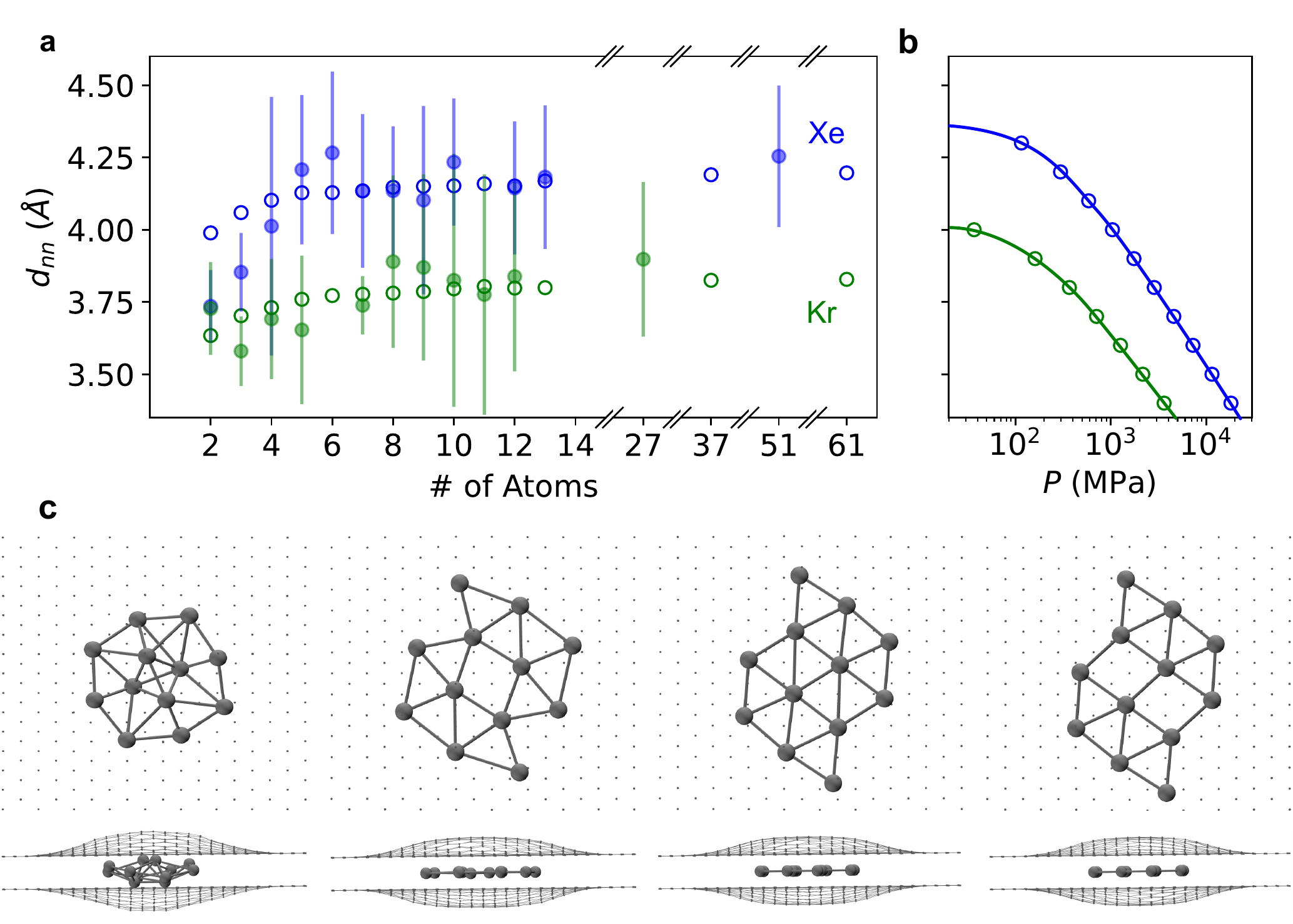}
	\caption{{\bf Interatomic distances and 3D to 2D transformation.}
        (a) Measured interatomic distances $d_{nn}$ (full circles) in the experimentally observed Kr (green) and Xe (blue) clusters as a function of the cluster size.
        Each data point is averaged over all images where the structure was clearly visible, taking always the mean value per image.
        The errorbars show the standard deviation of the data.
        Open symbols correspond to the simulation results.
        (b) Relationship between the pressure (plotted on logarithmic scale) exerted on the simulated 2D noble gas structure and the interatomic separation.
        The lines are guides for the eye.
        (c) Sequence of images showing selected steps from structural optimization starting with the twelve-atomic 3D Kr cluster in a graphene sandwich, which turns into a 2D structure.
	   The change in energy from left to right between the configurations is $-5.564$, $-0.565$ and $-0.033$~eV, respectively.
	}
	\label{fig:transition}
\end{figure}

Nearest neighbor distances for the clusters are shown in Fig.~\ref{fig:transition}a.
Both experimental and simulated results show clearly the same trend, where the smallest clusters have the shortest interatomic distances, with the strongest effect seen in the experimental data for the Xe clusters.
In all cases, the values are fairly constant with a slowly increasing trend for cluster sizes of $N\geq 10$.
Comparing $d_{nn}$ to the corresponding pressures estimated via simulations (Fig.~\ref{fig:transition}b) shows that the smallest clusters are under a pressure of up to 1~GPa, whereas for the larger clusters this drops to around 0.3~GPa.
These values are in the same range, but somewhat smaller than 1~GPa that has been estimated for 10~nm-sized clusters experimentally~\cite{vasu_van_2016}, and significantly lower than 30~GPa estimated for smaller He and Ar clusters computationally~\cite{villarreal_breakdown_2021}.
However, in neither of those cases were the clusters encapsulated between two graphene sheets.

The increasing pressure when the cluster size decreases is likely due to the higher necessary local curvature for graphene required for accommodating the smaller clusters.
Additionally, the pressure for the smallest experimental clusters appears to increase even faster than what is predicted by the simulations.
This can be related to the size of the point defect trapping the clusters, as discussed below.

To understand the influence of the graphene sandwich on the shape of the crystals, we carried out further structure optimization for cluster sizes up to 18 atoms (Fig.~S8).
These show, that in all cases, the optimum shape for the cluster (even at zero temperature) is three-dimensional for clusters larger than three atoms.
However, when the cluster is placed inside the graphene sandwich, it relaxes into a two-dimensional structure.
All clusters retain the 2D shape between the graphene sheets, even up to 61 atoms, which was the maximum simulated cluster size.
The change between the 3D and 2D structures is associated with an energy cost of ca. $60$~meV/atom for the Xe$_{18}$ cluster.
This energy is compensated by the interface energy between the cluster and graphene, and the lower deformation energy for the graphene sandwich in the flatter configuration.
An upper estimate for the interface energy (per atom) can be obtained by calculating the energy associated with an individual noble gas atom placed within two graphene sheets (total energy minus the energy of the similarly deformed graphene sheets without the additional atom).
This is ca. 149~meV for Kr and ca. 66~meV for Xe, which shows that the interface energy indeed has a similar order of magnitude as the energy difference of the 3D and 2D clusters.
Indeed, when graphene is removed from around the simulated 2D clusters, they transform to the 3D shape during further relaxation.

It is rather surprising, that the small clusters remain stationary at room temperature for long enough to allow their imaging in the microscope.
In fact, one could even expect that all noble gas atoms would escape through edges from between the two graphene layers before they are able to form solid structures.
While loosing all atoms is unlikely due to the large size of the samples (up to tens of $\mu$m for exfoliated samples and up to even mm for the CVD-grown ones) compared to atomic length scales, the motion of noble gases in the graphene sandwich warrants a discussion.
To study this, we turn to room temperature molecular dynamics simulations.
Our simulations of individual Kr and Xe atoms show that they are both highly mobile in a graphene sandwich with estimated speeds of ca. $316$~m/s and $233$~m/s, respectively.
For three-atom clusters this changes to ca. $334$~m/s and $231$~m/s  and for the Kr$_7$ cluster to ca. $369$~m/s at $300$~K.
(The corresponding speeds in gas under normal conditions, according to ideal gas law, would be $180$~m/s and $130$~m/s, respectively, for two dimensions).
Therefore, for all simulated clusters the speed within the graphene sandwich would be too high for imaging with the experimental time resolution of ca. 1~s per image.
The simulations further reveal that despite the weakness of the interaction between the noble gas atoms, already at $N=3$, all atoms within the clusters move together rather than separating (also seen for Kr$_7$ in Fig.~S6).
This is in good agreement with our experimental results, where we only rarely observe detachment of individual atoms from the clusters.

\begin{figure}[h!]
	\centering
	\includegraphics[width=1\textwidth]{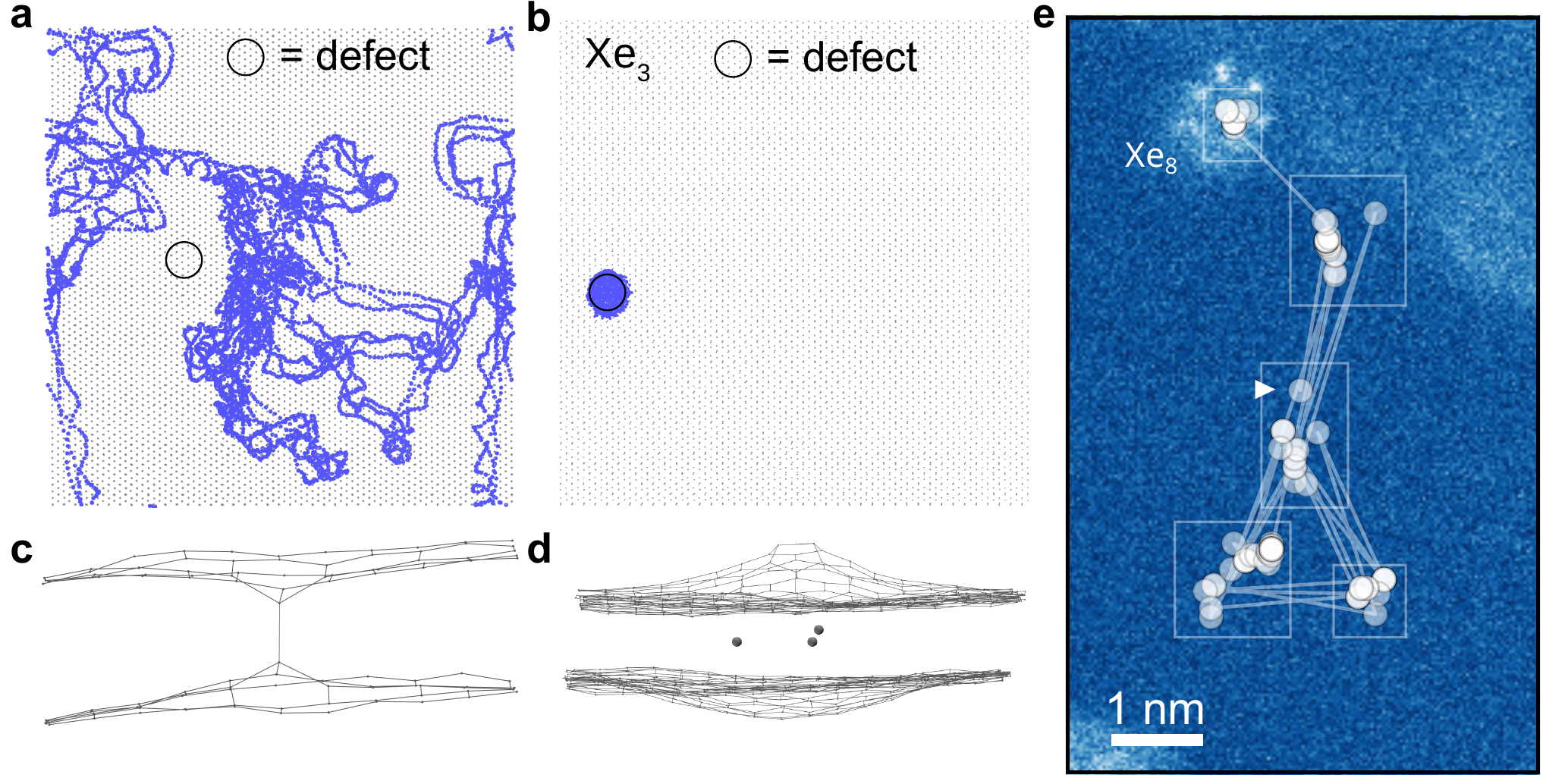}
    \caption{{\bf Cluster migration.}
        (a) Trajectory of a Xe$_3$ cluster in a graphene sandwich with an interlayer bond-defect, simulated for ca. $1$~ns via room temperature molecular dynamics.
        (b) Trajectory of a Xe$_3$ cluster in a graphene sandwich where one of the layers contains an inverse Stone-(Thrower)-Wales defect, simulated for ca. $1$~ns.
        For both simulations, the positions for the cluster atoms are shown for every $0.1$~ps, and the defect location is marked with an open circle.
        (c) Side view of graphene with the interlayer bond, where covalent bonds between the carbon atoms are shown with thin lines.
        (d) Side view of graphene with the inverse Stone-(Thrower)-Wales defect, and the Xe$_3$ cluster (shown with small balls).
        (e) Experimentally measured trajectory of an eight-atomic Xe cluster overlaid on a STEM-ADF image recorded at the end of the image sequence.
        The position of the cluster at the beginning of the sequence is marked with a white arrow, and the final position at the top of the image is visible due to the higher contrast of the Xe atoms.
        Positions for each of the images in the sequence are shown with semitransparent white circles, and the jumps between the images are marked with semitransparent white lines.
        The five distinct locations are highlighted with the rectangles.
        }
	\label{fig:dynamics}
\end{figure}

Obviously, regardless of the fast expected migration, the clusters remain static for long enough for us to be able to image them.
This suggests that the experimentally observed clusters are pinned to their positions either by corrugations associated with the surface contamination (many clusters appear at the edges of contamination) or by defects~\cite{lehtinen_effects_2010,trentino_atomic-level_2021}, some of which are expected to lead to out-of-plane distortion of graphene~\cite{kotakoski_atomic_2014}.
To test this hypothesis, we created two different defects in the graphene sandwich, and simulated the migration of a Xe$_3$ cluster in both cases.
We point out that these serve simply as examples of defects that would lead to either negative or positive Gaussian curvature in the graphene structure, and other defects may in reality be more likely to result from the low energy ion irradiation.
In the first case, we introduce an inter-layer covalent bond, which locally brings the two graphene sheets together (Fig.~\ref{fig:dynamics}a,c).
This leads to the cluster avoiding the defect position and instead migrating everywhere else.
In the second case, we introduced an inverse Stone-(Thrower)-Wales defect~\cite{lusk_nanoengineering_2008} that leads to negative curvature in graphene.
Here, the Xe$_3$ cluster is clearly attracted to the defect, and once trapped in its vicinity, it remains there (Fig.~\ref{fig:dynamics}b,d).
Although it can not be directly seen from the images, in both cases all atoms moved together throughout the whole simulation.
We assume that similar to these simulations, also in our experiments the observed clusters are attracted to defect sites.
Indeed, the ultralow-energy irradiation which is used to introduce the noble gas atoms into the graphene sandwich also by necessity creates defects into it.
Some of them in turn are associated with a negative curvature of the graphene sheet, which provides a favorable pocket for a small noble gas cluster.
Point defects are indirectly visible both in single and double layer parts of the sample in Fig.~S9a (by covalently bound heavier impurity atoms~\cite{inani_silicon_2019}), which additionally shows that noble gas clusters are only trapped in the double layer, highlighting that they can only be captured within the van der Waals gap between two graphene layers.
The size of the defect-related deformation may be a cause for the apparent unexpectedly high pressure for the smallest Xe clusters; the smallest clusters may fit into the deformation without further bending graphene by being squeezed tighter together leading to less bending in graphene but tighter pressed Xe atoms.

Interestingly, we also see occasionally that the clusters in the experiments jump between different sites in the graphene sandwich, even in areas void of visible contamination.
As an example, Fig.~\ref{fig:dynamics}e shows the trajectory of a Xe$_8$ cluster that was observed jumping between five different locations.
Another example of a jumping cluster is shown in Fig.~S10, where the cluster eventually merges with a stationary one.
We assume that the movement of the clusters is indirectly triggered by the imaging electron beam through scattering-related changes in the atomic configurations in the defects~\cite{kotakoski_stone-wales-type_2011} that serve as the pinning sites.
These pinning sites cannot arise from the moiré superstructure of the two graphene layers, because we observe similar behavior for samples regardless of the relative orientation of the graphene sheets (and even for mechanically exfoliated graphene with AB stacking).
Our interpretation is also in line with Ref.~\cite{cretu_migration_2010}, where individual W atoms were observed to similarly jump between discrete positions in graphene, assumed to be associated with point defects.
The main difference is that the W atoms migrate on top of graphene, whereas the noble gas clusters are confined within the van der Waals gap, which presumably has a significant influence on the associated energy barriers.

\begin{figure}[h!]	
	\centering
	\includegraphics[width=0.95\textwidth]{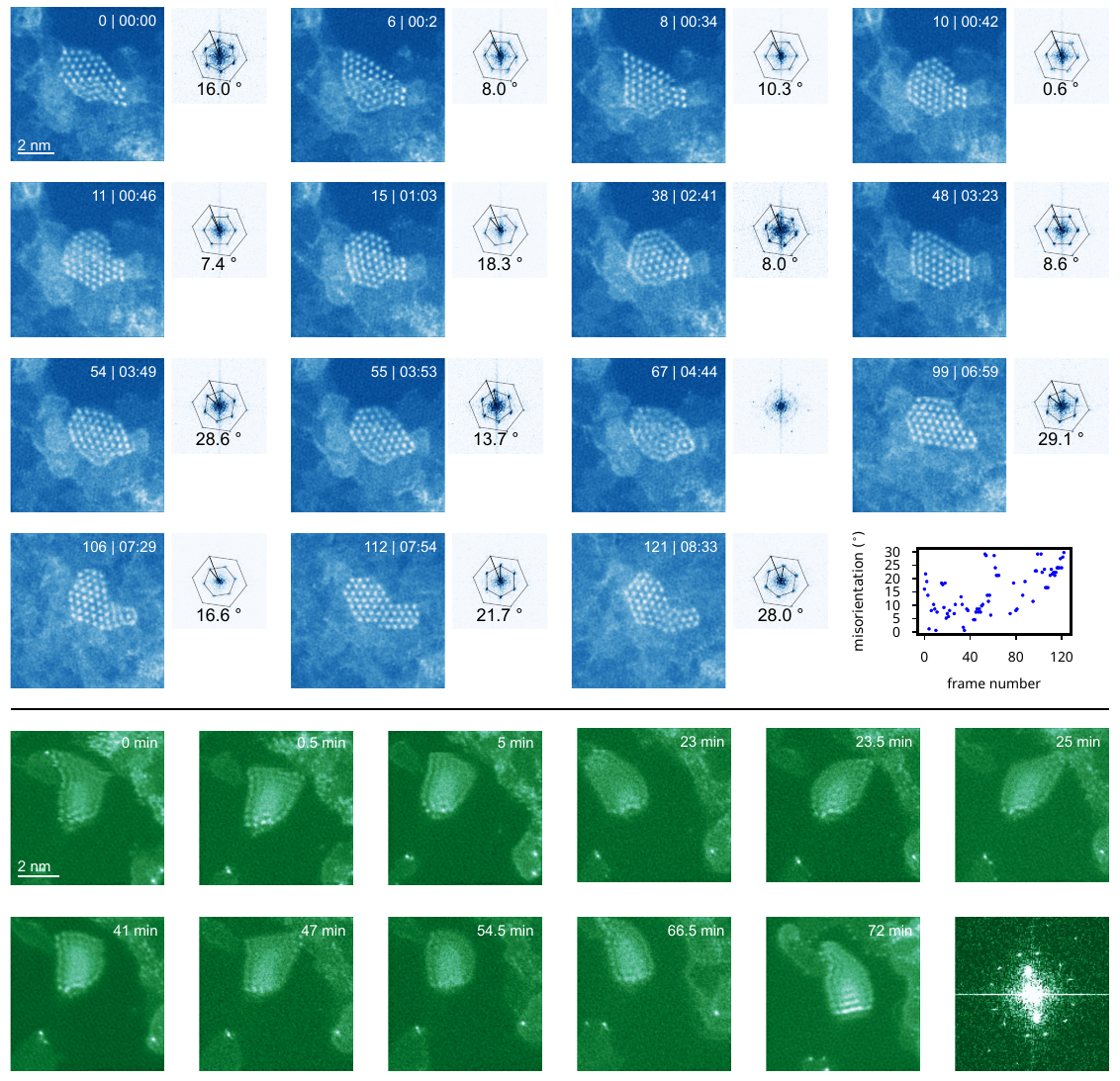}
    \caption{{\bf Structure of larger clusters.}
        (Top) Filtered STEM-MAADF images of a Xe$_{51}$ cluster within AB-stacked bilayer graphene that remains crystalline throughout the entire experiment (over $8$~min).
        The numbers on the images indicate the frame number within the image sequence and the corresponding time.
        The small images next to them show a fast Fourier transform of each image with overlaid hexagons highlighting the spots corresponding to graphene and to the noble gas cluster and the misorientation angle between them.
        These angles for each image frame are shown in the plot in the lower right corner.
        (Bottom) Filtered STEM-HAADF images of a Kr cluster of a similar size.
        The provided example of a fast Fourier transform shows the orientation of the two graphene sheets in the double layer sandwich (misorientation of 25$^\circ$).
        Due to the lack of crystallinity, the Kr cluster does not show characteristic spots.
        }
	\label{fig:big_crystals}
\end{figure}

One interesting question regarding the pressure-induced stability of 2D van der Waals atomic structures is under which conditions they remain solid.
The solid-fluid phase transitions carry information of the size-dependent interaction strength within the noble gas clusters, being of high relevance for future studies.
This is also interesting, since from a purely energetic point of view, one would expect smaller clusters to have a lower melting point and therefore to display liquid-like characteristics before the larger ones.
However, in the case of our clusters sandwiched between two graphene sheets, this is counterbalanced by the increasing pressure for the decreasing cluster sizes, as discussed above.
Nevertheless, it is natural to expect that there is a pressure- and temperature-dependant size limit, beyond which the energy cost of the undercoordinated atoms at the rim (corresponding to the surface energy of a 3D structure) becomes too large, and the structure undergoes a transformation into a three-dimensional shape.
Phase diagrams suggest a solid-to-liquid transformation at room temperature for pressures slightly above 1~GPa for Kr and around 0.5~GPa for Xe~\cite{david_a_young_phase_1991}.
Therefore, one would expect larger Kr clusters to loose their solid shape, which we indeed observe (Fig.~\ref{fig:big_crystals}).
Although according to simulations, both Xe$_{61}$ and Kr$_{61}$ clusters should retain the 2D shape, experimental images (Fig.~\ref{fig:big_crystals}) show a brighter area within the Kr structure, which may indicate that the structure is not strictly two-dimensional.
Indeed, Kr structures larger than $N \sim 16$ already start to show less defined atomic positions (at the rim already earlier, see Fig.~\ref{fig:crystals}a), and shapes that resemble liquid more than small crystallites.
However, Xe clusters remain two-dimensional and exhibit characteristics of solid crystallite structures (clearly discernible atomic positions with long-range order) up to much larger sizes, although the estimated pressure is only 0.3~GPa.
In fact, as shown in Fig.~S9b, the Xe clusters start to loose their solid characteristics only at the size of $N\sim 100$ atoms.

As is evident in all images of clearly liquid-like structures (Fig.~\ref{fig:big_crystals}, Fig.~S9b), the solid-to-fluid transition is also associated with loosing the strict two-dimensionality.
Although the energy per atom decreases faster for 3D clusters than for 2D ones (Fig.~S8), this energy gain is countered with a cost for larger deformation in graphene, which is larger for Xe than for Kr due to its larger size.
This may provide a partial explanation for why Xe remains solid for much larger structures than Kr.
Finally, we point out that as can be seen in Fig.~\ref{fig:big_crystals}, the larger clusters rotate during imaging, ruling out epitaxy with the graphene lattice.

Creation of small solid state noble gas clusters in a graphene sandwich, as was demonstrated here, opens the so-far unexplored frontier of encapsulated 2D van der Waals atomic solids.
Simulations show that the small clusters would assume a 3D shape without the graphene encapsulation, even at zero temperature, and that noble gas clusters are extremely mobile in the graphene sandwich, unless pinned to a position due to a deformation in at least one of the encapsulating graphene layers.
Although small clusters remain solid for both Kr and Xe, larger Kr clusters ($N>16$) start loosing their solid structure whereas Xe clusters remain solid up to $N\sim 100$.
Larger solid structures may be achieved by providing more structural rigidity for example through increasing the number of the encapsulating layers, which can be expected to increase the pressure.
Combined with {\it in situ} techniques for example allowing experimentation from cryogenic to elevated temperatures, encapsulated 2D noble gas structures provide exciting possibilities for studies in fundamental condensed matter physics ranging from research on the growth and atomic-scale dynamics in solids to phase transitions and topological defects.
They may also lead to well-defined systems for example for quantum information research.

\section*{Online content}

Supplementary material includes description of methods, overview images and electron energy loss spectra, a schematic presentation of the observed atomic structures, images of additional observed shapes of small Kr and Xe clusters, time series of a Kr cluster changing size over time, simulations of shape changes of a cluster, all different simulated 2D and 3D shapes, simulation results for the energetic differences between 2D and 3D clusters and an additional example of an image series of a Xe cluster that jumps within the graphene sandwich.

\bibliography{scibib}
\bibliographystyle{naturemag_doi_jk}

\clearpage

\subsection*{Methods}

\noindent {\bf Samples.} Few-layer graphene samples were fabricated using both mechanical exfoliation as well as stacking single-layer graphene grown via CVD (EasyTransfer graphene by Graphenea, Inc.) on top of each other.
Mechanically exfoliated samples were transferred using the typical transfer procedure \cite{meyer_hydrocarbon_2008} to Quantifoil Holey Carbon TEM grids with a hole size for the suspended sample of either 1.2 or 2.0~$\mu$m, depending on the sample.
CVD-samples were transferred to the same grids following the manufacturer instructions (two consecutive transfers onto the same grid were carried out to create the necessary double layer).
\\

\noindent {\bf Ion irradiation.} Most samples were irradiated at the University Helsinki using the $500$~kV ion implanter KIIA operated at $20$~kV with an electrostatic deceleration lens slowing the ions down to $20-100$~eV.
To reduce sample contamination during the implantation process, some of the samples were baked in air for $1$~h at $300^\circ$C and inserted warm into the irradiation setup.
For Kr-irradiated samples, successful implantation between two graphene layers was observed at 60~eV irradiation, and for Xe-irradiated samples at $55$~eV and $65$~eV.

Additional irradiations were carried out within the vacuum system CANVAS~\cite{mangler_materials_2022} at the University of Vienna, which connects the Nion UltraSTEM 100 with other experimental stations.
Here, a plasma source was used to create a Xe beam.
The energy of the plasma was measured by biasing a Faraday cup and measuring the current as a function of the bias voltage.
For implantation, the beam was decelerated using sample bias until the energy of the ions was ca. 30~eV.
Results from these irradiations are shown in Fig.~S9, whereas all other results are from samples irradiated at the University of Helsinki.
\\

\noindent {\bf Electron microscopy and spectroscopy.} All samples were imaged at the University of Vienna with the Nion UltraSTEM 100 dedicated scanning transmission electron microscopy (STEM) instrument at $60$~kV using either the high or the medium angle annular dark field (HAADF and MAADF) detector with annular ranges of $60-200$~mrad and $80-300$~mrad, respectively.
The convergence semi-angle was $30$~mrad, and a typical beam current ca. $30$~pA.
Electron energy loss spectroscopy was carried out with the same instrument with a Gatan PEELS~666 spectrometer retrofitted with an Andor iXon 897 electron-multiplying charge-coupled device camera using a collection semi-angle of ca. $35$~mrad.
Before being inserted into the vacuum system connected to the microscope, the samples were baked in vacuum at ca. 150$^\circ$C for at least 8~h.
Some samples were additionally cleaned before imaging with a laser at the microscope column ($50$~$\mu$s laser pulse, power $20$~mW, wavelength $473$~nm and spot size ca. $700$~$\mu$m$^2$).
The images were calibrated using the auto scaling algorithm~\cite{madsen_fourier-scale-calibration_2022} based on the lattice spacing of graphene, and the positions of the noble gas atoms were detected with the automatic blob detector~\cite{van_der_walt_scikit-image_2014}.
The samples remained at room temperature during imaging.
Due to the excellent conductive properties of graphene, no charging of graphene or the sandwiched structures took place during imaging.
\\

\noindent {\bf Image processing.} Image filtering was done by applying a double Gaussian~\cite{krivanek_atom-by-atom_2010} and a Fourier filter.
For the double Gaussian filter, the inner radius was set as 0.15 times the distance from the centre to the first set of graphene spots in the fast Fourier transform (FFT) and the radius of the outer Gaussian was set as 1.3 times the respective distance to the outer set of FFT spots.
In the Fourier filter, the spots corresponding to graphene were deleted and thus lattice was removed after back-transformation.
\\

\noindent {\bf Atomistic simulations.} The molecular dynamics simulations were conducted using the Large-scale Atomic/Molecular Massively Parallel Simulator (LAMMPS).
The periodic system consisted of two $10\times 10$~nm$^{2}$ AB-stacked graphene layers (7872 C atoms) with the interlayer distance of $3.39$~{\AA} relaxed to a local energy minimum with the Polak-Ribiere modified conjugate gradient algorithm~\cite{notay_flexible_2000}.

The defected graphene sheets with an interlayer bond and a defect with negative curvature were created by introducing the defect to the relaxed AB-stacked graphene and re-relaxing the system. 
After the initial relaxations, the noble gas atoms were introduced between the graphene layers in the $2$D form as observed in the experiments and the system was relaxed again.

Optimised $3$D clusters were obtained from these $2$D forms by annealing the cluster without graphene at $80$~K over $40$~ps followed by cooling down to $1$~K during $800$~ps and relaxation. 
The rapid annealing and slow cooling was introduced to escape possible local energy minima.

For observing the cluster transformation into $2$D, two optimised $3$D clusters (Kr$_6$ and Kr$_{12}$) were individually placed between the relaxed graphene sheets and then relaxed again. 

The room temperature dynamics of a single atom, a three-atom cluster and a seven-atom cluster were simulated by first ramping the temperature from $0$~K to $300$~K over $0.2$~ns, after which the simulation was run at $300$~K for another $1$~ns.
The output was printed every $500$ steps with a time step of $0.2$~fs. 

The pressure of the Kr and Xe clusters were calculated from a $4\times5$ atom semiperiodic cell.
The bond lengths were scaled to the selected value and the forces ($F_y$) between atoms taken from the initial structure along the nonperiodic direction when conjugent gradient energy minimisation was started.
Pressure was calculated with $P_y=F_y/A$, where $A$ is the 2D area given by the distance between two neighboring atoms and the height of the van der Waals gap containing the cluster ($3.4$~Å for Kr, $3.3$~Å for Xe).

Kr and Xe interactions were described via the Lennard-Jones potential, with the Kr-Kr and Xe-Xe parameters $\varepsilon_\mathrm{Kr} = 30.7$~meV, $\sigma_\mathrm{Kr} = 3.6233$~{\AA} and $\varepsilon_\mathrm{Xe} = 44.1$~meV, $\sigma_\mathrm{Xe} = 3.9460$~{\AA})~\cite{rutkai_how_2017} and the Kr-C and Xe-C with parameters ($\varepsilon_\mathrm{Kr-C} = 10.7$~meV, $\sigma_\mathrm{Kr-C} = 3.5116$~{\AA} and $\varepsilon_\mathrm{Xe-C} = 12.8$~eV, $\sigma_\mathrm{Xe-C} = 3.673$~{\AA}) calculated using the Lorenz-Berthelot mixing rules~\cite{lorentz_ueber_1881,bethelot_mixing_1898}.
Carbon-carbon interaction was described by the AIREBO potential~\cite{stuart_reactive_2000}, including the van der Waals interaction.

\section*{Acknowledgments}

{\bf Funding:} Funding through the Austrian Science Fund (FWF) within projects P31605, P34797 and M2596, and generous grants for computational resources from the Vienna Scientific Cluster are gratefully acknowledged.

\section*{Author contributions}

M.L. and J.K. designed the experiments. M.L. prepared the samples and carried out the ion irradiation with the help of K.M. and H.{\AA}.
H.{\AA}. carried out the molecular dynamics simulations and analysed that data together with J.K.
M.L. and A.T. carried out the microscopy with the help from C.M., K.M. and J.K. M.L. and J.K. analyzed the experimental data, plotted the figures, and wrote the first draft.
All authors were involved in writing the manuscript.
J.K. supervised the project.

\section*{Competing interests}

The authors declare no competing interests.

\section*{Data and materials availability}

All original data used in producing the presented results is available in the supplementary material and through the University of Vienna repository Phaidra~\cite{langle_original_nodate}.

\clearpage

\section*{Supplementary Material}

\setcounter{figure}{0}
\makeatletter 
\renewcommand{\thefigure}{S\@arabic\c@figure}
\makeatother

\begin{figure}[h!]	
	\centering
	\includegraphics[width=1\textwidth]{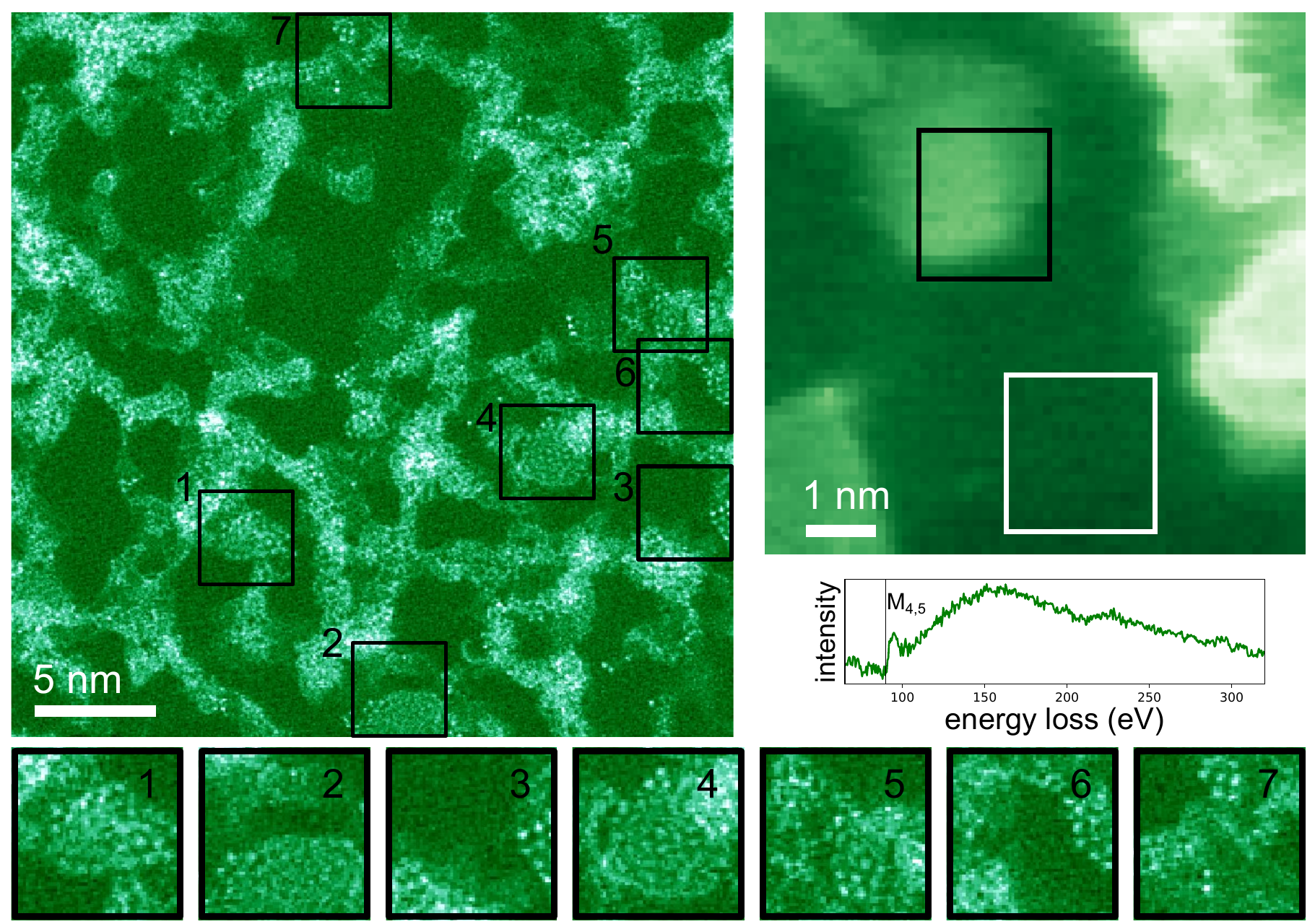}
	\caption{{STEM-MAADF overview image of an area containing several Kr clusters (some of which are highlighted by numbered rectangles with magnified versions shown at the bottom) and spectral map.}
		The electron energy loss spectral map was recorded over a cluster with approximately $50$-$70$ atoms, shown on the bottom of Fig.~4.
		The contrast here corresponds to the integrated spectrum for each pixel.
		The signal from the Kr $M_{4,5}$ core loss peak arises from the area spatially located at the cluster, as can be seen from the spectrum obtained by subtracting the graphene signal (from within the area marked by the white rectangle) from the signal corresponding to the area marked by the black rectangle (the spectra were normalised to the values before the peak onset at 89~eV).
	}
	\label{fig:spectral_map_Kr}
\end{figure}

\begin{figure}[h!]	
	\centering
	\includegraphics[width=1\textwidth]{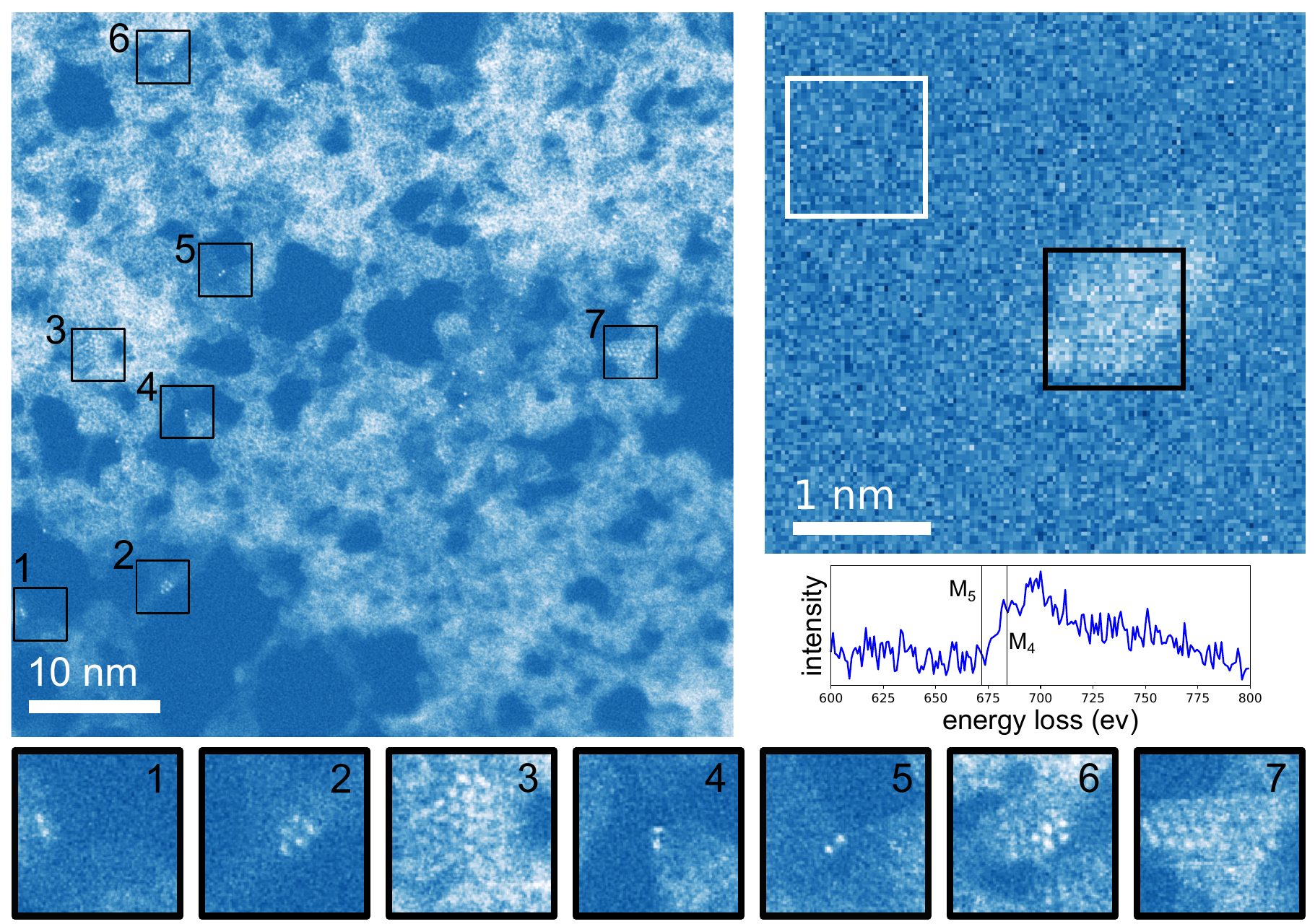}
	\caption{{STEM-MAADF overview image of an area containing several Xe clusters (some of which are highlighted by numbered rectangles with magnified versions shown at the bottom) and spectral map.}
		The electron energy loss spectral map was recorded over a Xe$_6$ cluster (same cluster as Xe$_6$ in Fig.~1 of the main article).
		The contrast here corresponds to the integrated spectrum for each pixel.
		The Xe $M_{4}$ and $M_{5}$ core loss peaks arise from the area spatially located at the cluster, as can be seen from the spectrum obtained by subtracting the graphene signal (from within the area marked with a white rectangle) from the signal corresponding to the area marked by the black rectangle (the spectra were normalized to the values before the peak onset at 672~eV.
        Two very bright pixels, possibly caused by cosmic rays hitting the detector, were replaced by their average surrounding.
	}
	\label{fig:spectral_map_Xe}
\end{figure}

\begin{figure}[h!]	
	\centering
	\includegraphics[width=0.4\textwidth]{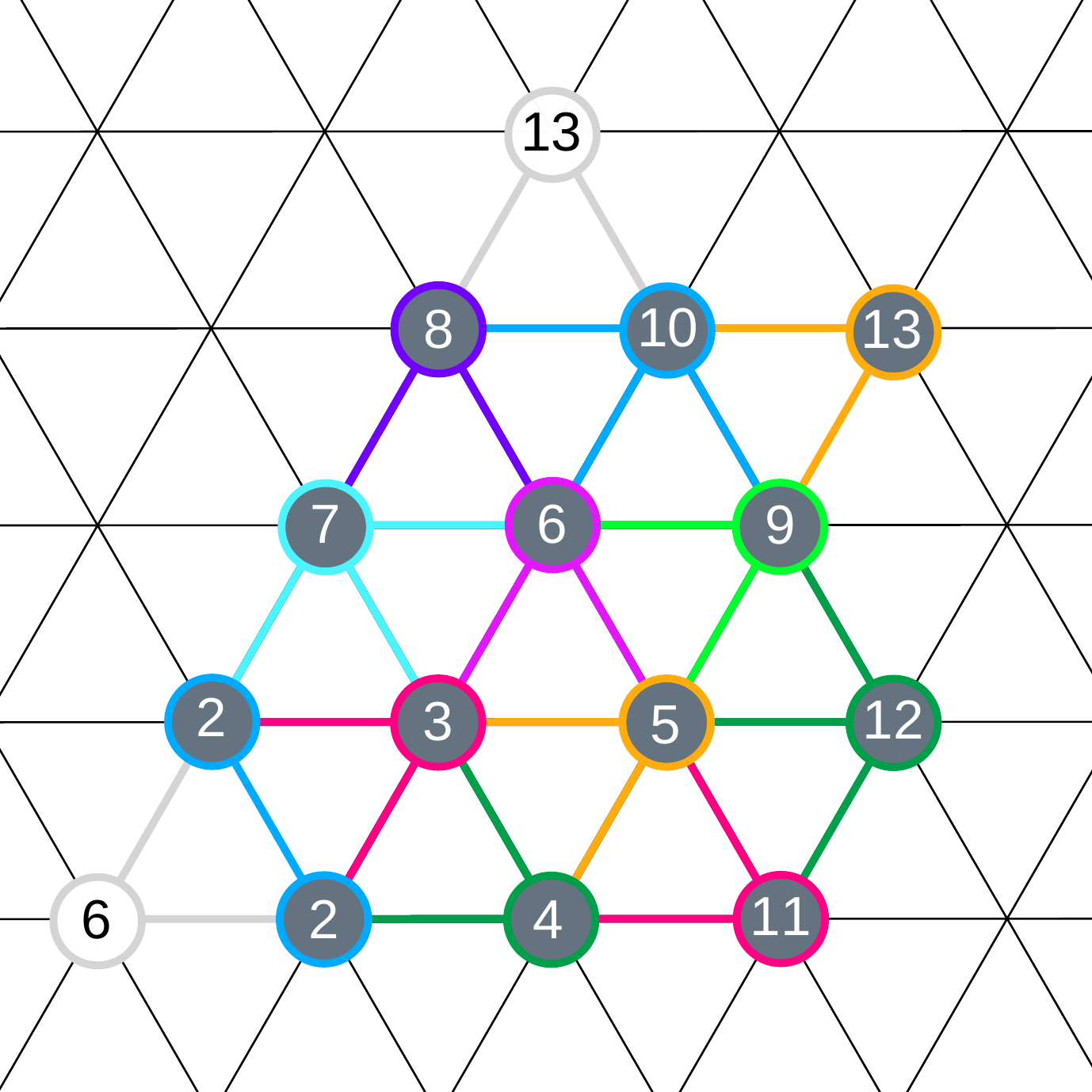}
    \caption{{Schematic presentation of the observed atomic structures.}
    The numbers indicate the size of the cluster (positions marked with $i\leq N$ constitute the cluster with $N$ atoms).
    For each added atom, the bonds with the same color as the circle are new compared to the cluster with one less atom.
    The position shown with gray outline and black number $6$ corresponds to the position of the sixth atom in Xe$_6$ in the observed cluster, whereas the colored circle with white $6$ is the position of the sixth atom in all clusters with $N>6$.
    In contrast, the colored circle with white $13$ corresponds to the observed structure of Xe$_{13}$, whereas the gray circle with black $13$ would lead to a more symmetric structure but with the same number of nearest neighbor interactions.
    }
	\label{fig:structures}
\end{figure}

\begin{figure}[h!]	
	\centering
	\includegraphics[width=1\textwidth]{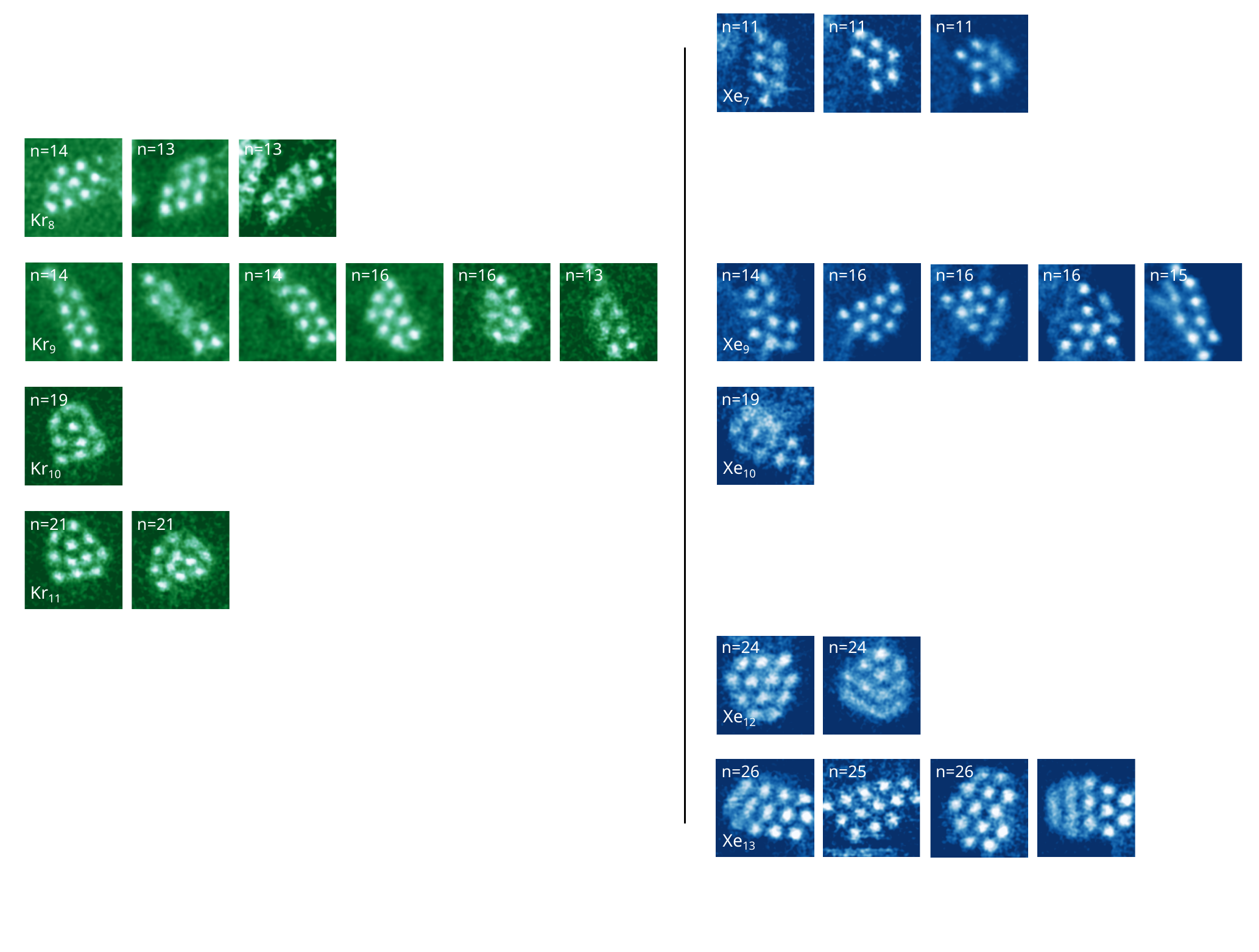}
    \caption{{Filtered STEM-ADF images (see Methods) of small Kr and Xe clusters displaying all additionally observed shapes in our experiments.}
    All images have an area of $2\times 2$~nm$^2$. Left are Kr and right are Xe clusters. The overlaid values for $n$ show the number of nearest-neighbor pairs for each configuration.
    }
	\label{fig:configurations}
\end{figure}

\begin{figure}[h!]	
	\centering
	\includegraphics[width=0.9\textwidth]{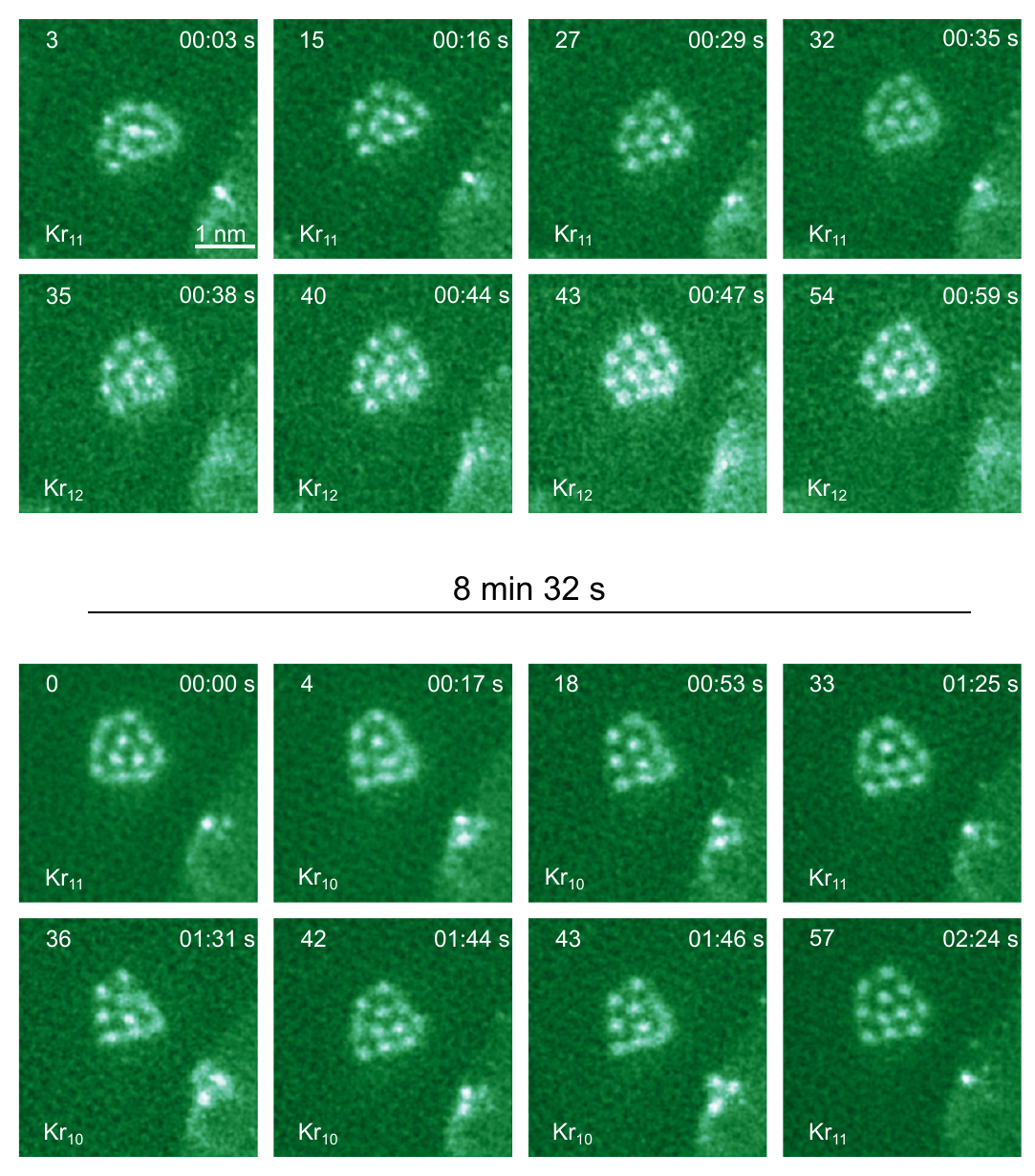}
    \caption{{Filtered STEM-HAADF images (see Methods) of a Kr cluster changing size over time.} The labels correspond to the frame numbers within the image sequence and the corresponding time.
    The second images series was recorded 8~min 32~s after the first frame of the first sequence.
	}
	\label{fig:cluster_growing}
\end{figure}

\begin{figure}[h!]	
	\centering
	\includegraphics[width=1\textwidth]{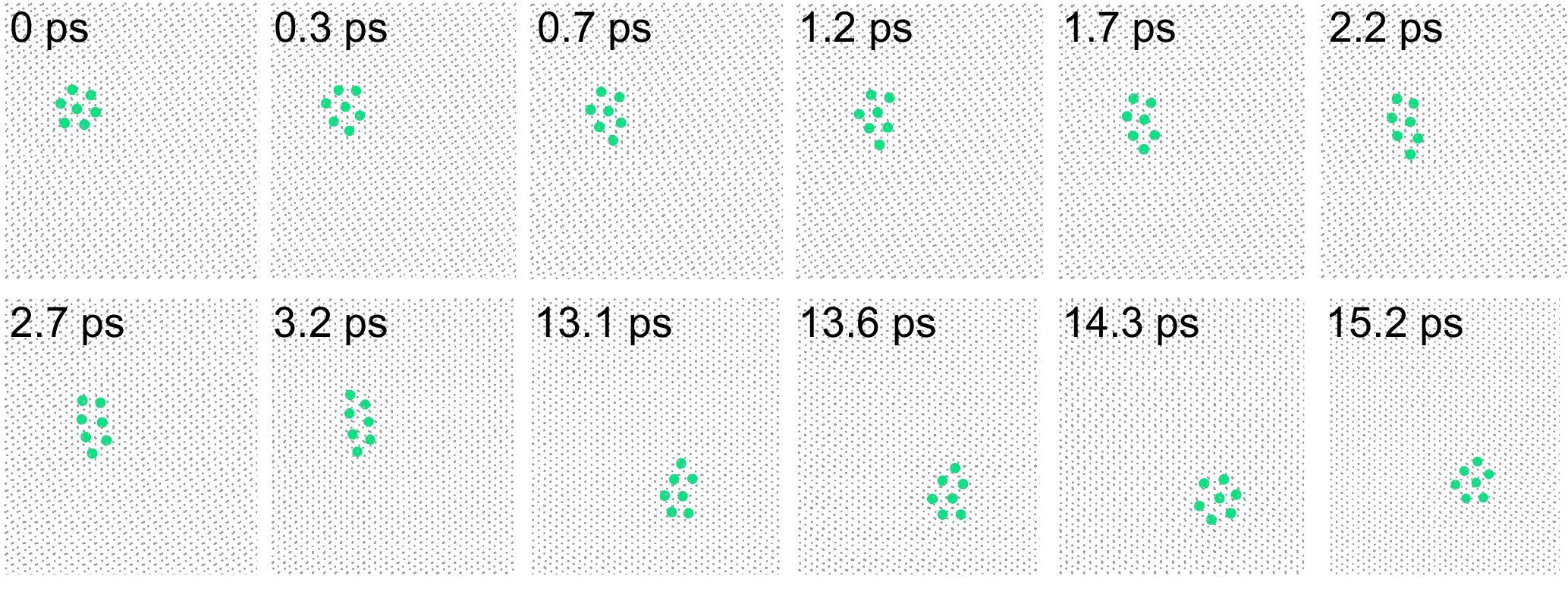}
	\caption{Snapshots of a room temperature molecular dynamics simulation showing configuration changes for a Kr$_7$ cluster.
		}
	\label{fig:sim_clustershape}
\end{figure}

\begin{figure}[h!]	
	\centering
	\includegraphics[width=1\textwidth]{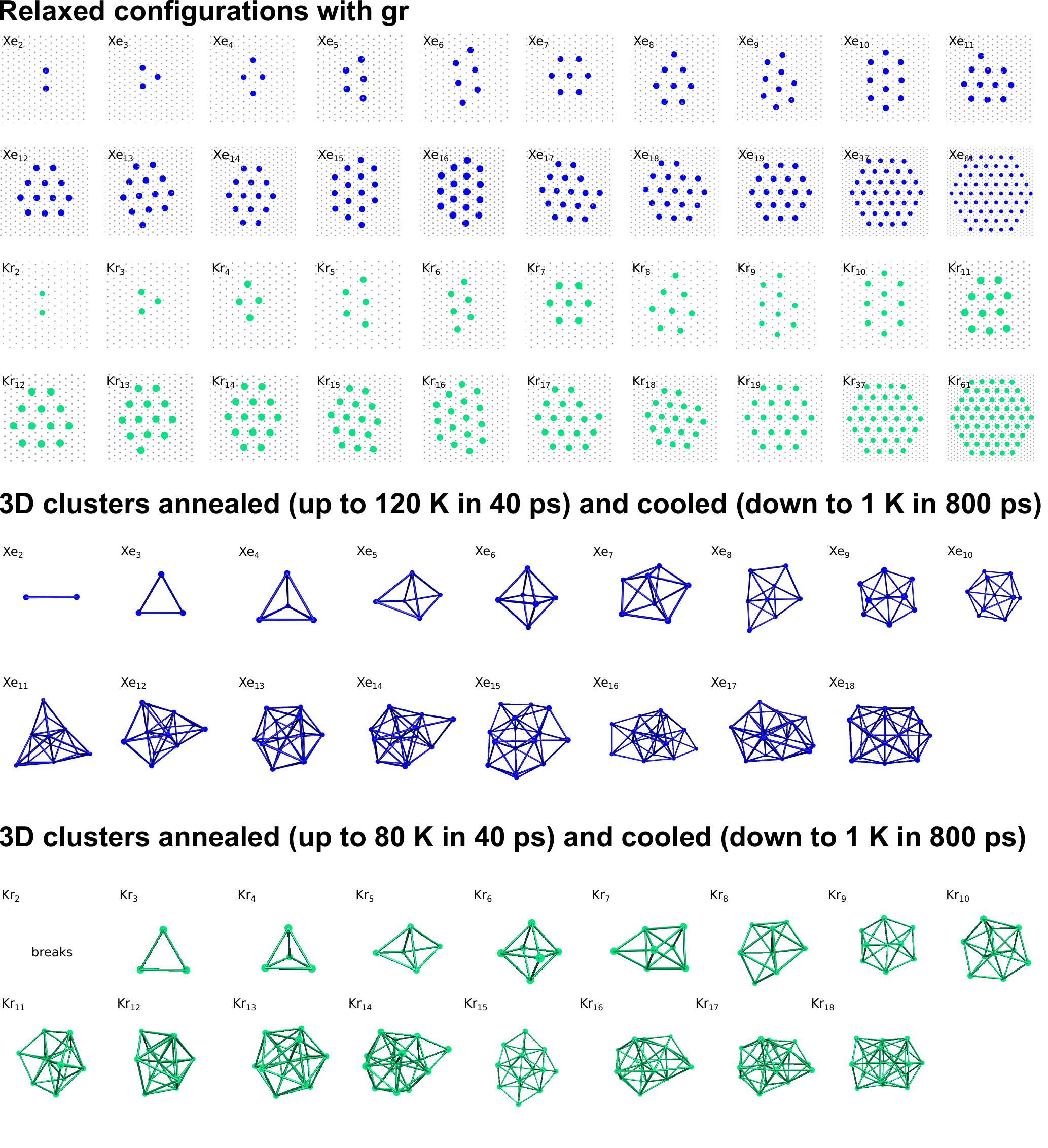}
	\caption{{Lowest-energy configurations of simulated cluster shapes in 2D and 3D.}
	}
	\label{fig:simulated_shapes}
\end{figure}

\begin{figure}[h!]	
	\centering
	\includegraphics[width=1\textwidth]{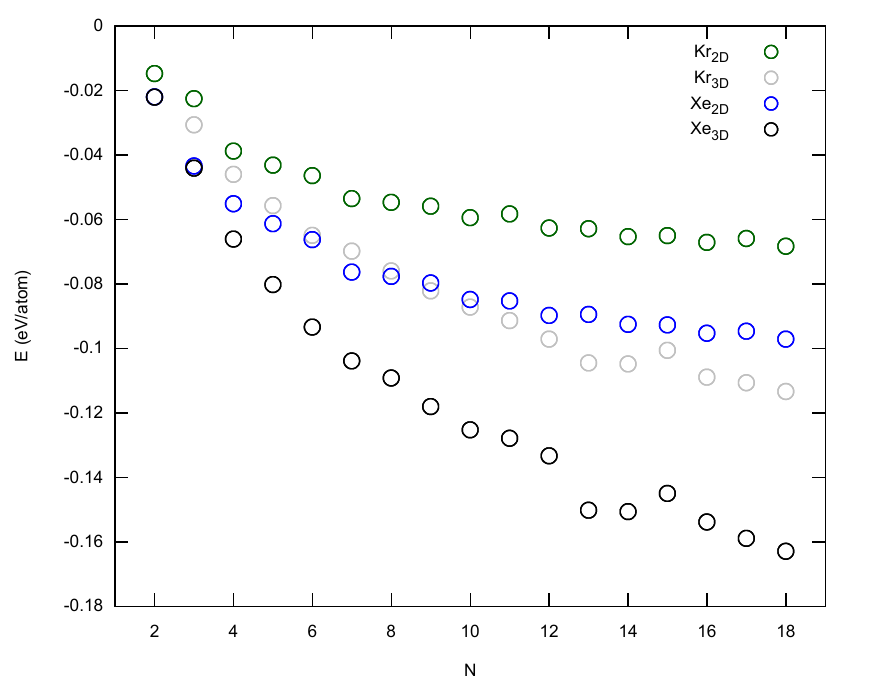}
	\caption{{Energy difference between 2D and 3D clusters.}
		The lowest-energy 3D structures were obtained by removing the graphene encapsulation and relaxing the clusters again after annealing.
	}
	\label{fig:energy_2D_3D}
\end{figure}

\begin{figure}[h!]	
	\centering
	\includegraphics[width=1\textwidth]{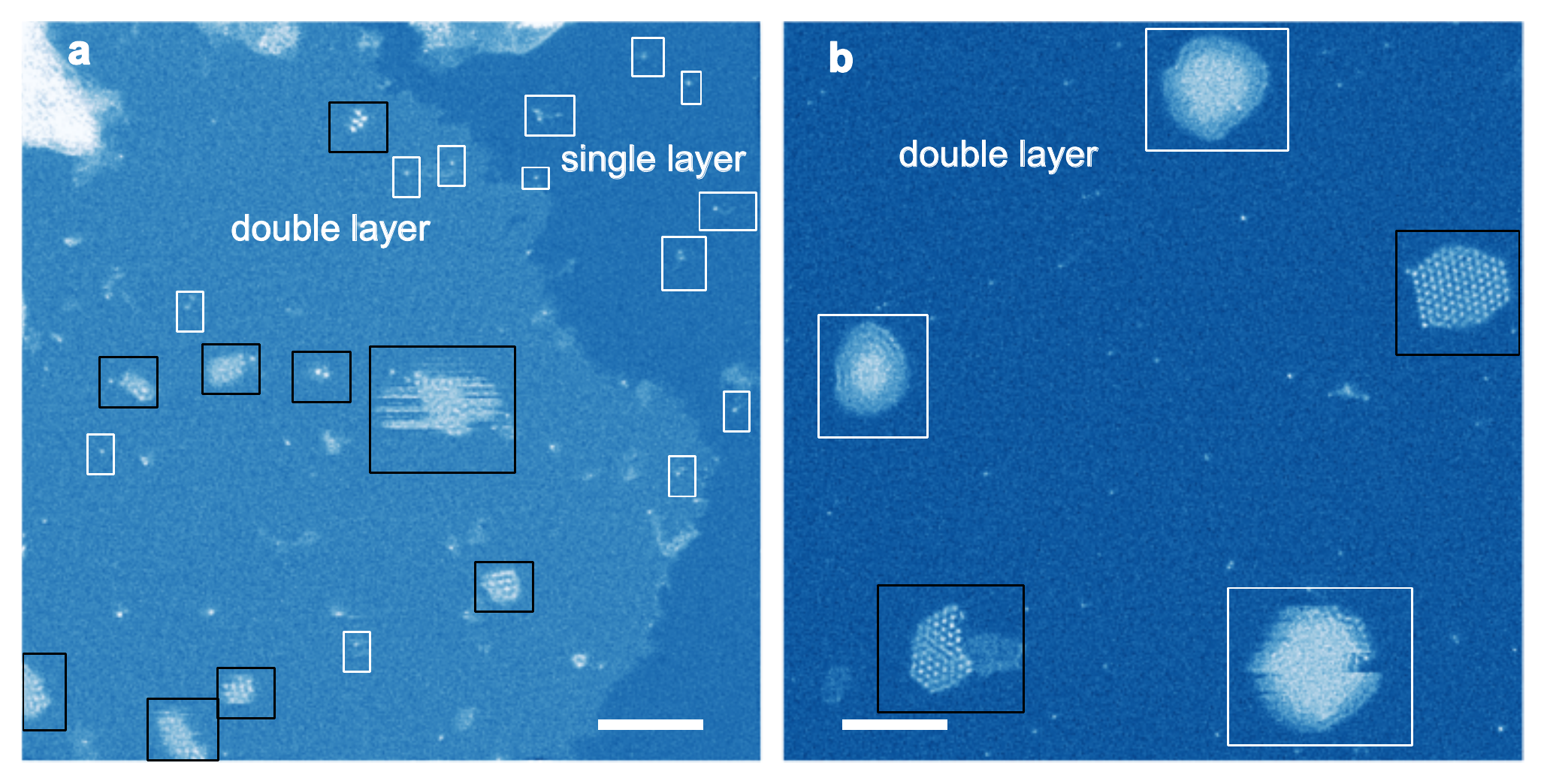}
    \caption{{Filtered STEM-MAADF images (see Methods) showing (a) a step edge between a double layer and a single layer and (b) Xe clusters with $N\approx 100$ implanted in a double layer.}
            In panel (a) white rectangles mark some positions where Si heteroatoms have been trapped at defects in both areas, whereas black rectangles show locations of noble gas clusters, that are exclusively in the double layer area.
            In panel (b), black rectangles show solid Xe clusters and the white ones mark clusters that have a liquid-like structure.
            The larger solid cluster has $N = 93$ atoms, whereas the smallest liquid-like structure has only a few atoms more.
			Both images have been recorded on the same sample prepared from CVD graphene, which--in contrast to all other samples--was implanted with Xe clusters with a plasma source at the University of Vienna, as described in Methods.
			Both scale bars are 5~nm.}
	\label{fig:merging_clusters}
\end{figure}

\begin{figure}[h!]	
	\centering
	\includegraphics[width=1\textwidth]{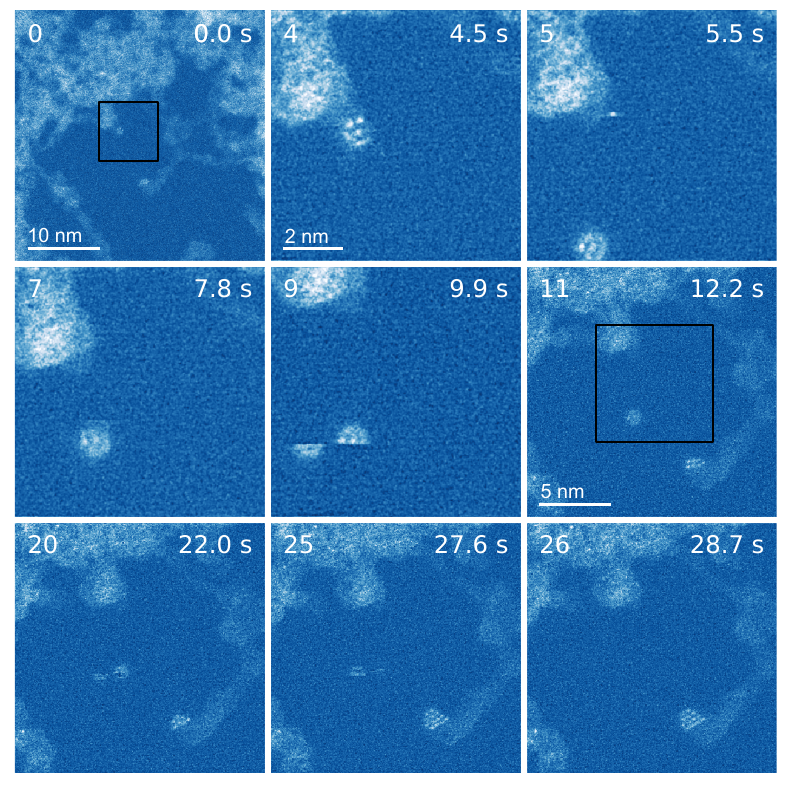}
    \caption{{Filtered STEM-MAADF images (see Methods) of a Xe$_6$ cluster jumping between different positions, and finally merging with another cluster.}
		The labels correspond to the frame numbers within the image sequence and the corresponding time.
        The black squares on frames $0$ and $11$ mark the area shown in frames $4-9$ and $20-26$.}
	\label{fig:merging_clusters}
\end{figure}

\end{document}